\documentstyle[mnextra]{mn} 
\input{epsf}
\def \etal {et al.\ }

\nobrackets     

\newcommand{\galf}{\textsc{galform}}

\begin{document}

\title[Gas dynamics in galaxy formation models]{A Comparison of gas dynamics in SPH and semi-analytic models of galaxy formation}

\author[J. C. Helly et al.] {
\parbox[h]{\textwidth}{John C. Helly$^1$, Shaun Cole$^1$, Carlos S. Frenk$^1$,
Carlton M. Baugh$^1$, Andrew Benson$^2$, Cedric Lacey$^1$, Frazer R. Pearce$^3$}
\vspace*{6pt} \\ 
$^1$Department of Physics, University of Durham, Science Laboratories, South 
Road, Durham DH1 3LE, United Kingdom \\
$^2$California Institute of Technology, MC105-24, Pasadena CA 91125, USA \\
$^3$Physics and Astronomy, University of Nottingham, Nottingham, NG7 2RD\\
}
\maketitle
 
\begin{abstract}
We compare the results of two techniques used to calculate the
evolution of cooling gas during galaxy formation: Smooth Particle
Hydrodynamics (SPH) simulations and semi-analytic modelling. We
improve upon the earlier statistical comparison of Benson et al. by
taking halo merger histories from the dark matter component of the SPH
simulation, which allows us to compare the evolution of galaxies on an
object-by-object basis in the two treatments. We use a
``stripped-down'' version of the semi-analytic model described by
Helly \etal which includes only shock heating and
radiative cooling of gas and which is adjusted to mimic the resolution
and other parameters of a comparison SPH simulation as closely as
possible. We compare the total mass of gas that cools in halos of
different mass as a function of redshift as well as the masses and
spatial distribution of individual ``galaxies.'' At redshift $z=0$,
the cooled gas mass in well-resolved halos agrees remarkably well (to
better than $\sim$20\%) in the SPH simulation and stripped-down
semi-analytic model.  At high redshift, resolution effects in the
simulation become increasingly important and, as a result, more gas
tends to cool in low mass halos in the SPH simulation than in the
semi-analytic model. The cold gas mass function of individual galaxies
in the two treatments at $z=0$ also agrees very well and, when the
effects of mergers are accounted for, the masses of individual
galaxies and their 2-point correlation functions are also in excellent
agreement in the two treatments. Thus, our comparison confirms and
extends the earlier conclusion of Benson et al. that SPH simulations
and semi-analytic models give consistent results for the evolution of
cooling galactic gas.
\end{abstract}
\begin{keywords}
galaxies: formation - methods: numerical
\end{keywords}

\section{Introduction}
\label{sec:intro}

A range of physical processes are responsible for the formation and evolution of the galaxies we see in the universe today. The starting point for current hierarchical cold dark matter models of galaxy formation is the gravitational amplification and eventual collapse of primordial density fluctuations to form the dark matter halos in which stars and galaxies may form. This process is now quite well understood, and predictions of halo mass functions from analytic techniques such as Press-Schechter theory (Press \& Schechter \shortcite{ps74}) and its extensions (\cite{bond91,bower91,lc93,st01}) are in good agreement with numerical simulations (e.g. \cite{gross98,governato99,j2001}).

Unfortunately, the behaviour of the baryonic component of the universe is more complex and less well understood. While the dynamics of the dark matter are determined by gravitational forces alone, gas is subject to hydrodynamical forces and radiative effects. The situation is further complicated by the absence of a complete theory of star formation and the fact that star formation involves length and mass scales many orders of magnitude smaller than the galaxies themselves forces those modelling galaxy formation to resort to recipes and prescriptions to obtain star formation rates. Nevertheless, semi-analytic models have met with considerable success, for example in reproducing the local field galaxy luminosity function and distributions of colour and morphology (e.g. \cite{cole91}; \cite{cole94}, \shortcite{cole2k}; \cite{wf91,ls91,sp99}) and galaxy clustering properties (e.g. \cite{kauffmann99,b2000,wechsler2001}). In this work, we compare two possible ways of modelling the process which provides the raw material for star formation -- the cooling of gas within dark matter halos. Such a model is a necessary part of almost any treatment of the hierarchical formation of galaxies, yet there is still some uncertainty as to which of the approaches currently in use are reliable and whether they are in good agreement.

While Eulerian numerical techniques may be employed in the modelling of galaxy formation in cosmological volumes (e.g. \cite{co2k}), here we concentrate on the Lagrangian method known as smoothed particle hydrodynamics (SPH), first described by Lucy (\shortcite{lucy77}) and Gingold \& Monaghan (\shortcite{gm77}). SPH simulations have been able to predict the formation of objects of approximately galactic mass with appropriate abundances in a cosmological context (e.g. \cite{khw92,nw93,esd94,sm95,kwh96,frenk96,ns99}; Pearce \etal \shortcite{pearce99}, \shortcite{p2001}) and allow the investigation of the dynamics of galaxies within clusters and the spatial distribution of galaxies. 

Semi-analytic and SPH galaxy formation models rely on very different sets of assumptions and approximations. For example, semi-analytic models assume that dark matter halos are spherically symmetric and that infalling gas is shock-heated to the virial temperature of the halo, whereas SPH simulations impose no restrictions on halo geometry but assume that continuous distributions of gas and dark matter may be well represented by a limited number of discrete particles. Consequently, SPH and semi-analytic models have complementary strengths and weaknesses. Semi-analytic models are computationally much cheaper than simulations, which allows extremely high mass resolution in halo merger trees and more thorough investigation of the effects of varying parameters or the treatment of particular processes. SPH simulations contain fewer simplifying assumptions but have limited dynamic range and without sufficiently large numbers of particles may suffer from numerical effects.

The aim of this paper is to compare SPH and semi-analytic treatments of the gas dynamics involved in galaxy formation in order to gauge the effects of the uncertainties present in the two techniques. A previous comparison carried out by Benson \etal (\shortcite{b2001}) found that SPH and semi-analytic models give similar results for the thermodynamic evolution of cooling gas in cosmological volumes. In particular, the global fractions of hot gas, cold dense gas and uncollapsed gas agreed to within 25\% and the mass of gas in galaxies in the most massive halos differed by no more than 50\%. However, their analysis was restricted to a statistical comparison because their semi-analytic model employed merger histories created using a Monte-Carlo algorithm, that of Cole \etal (\shortcite{cole2k}). We improve on the work of Benson \etal by calculating the merger trees directly from the simulations so that the merger histories of the halos in the semi-analytic and SPH treatments are the same. This removes a source of uncertainty from the comparison, since any differences between the models must be due entirely to differences in the treatment of the \emph{baryonic} component. Our method also allows a comparison between halos on an individual basis and lets us investigate whether the dependence of the cold gas mass on the halo's merger history is the same in the SPH and semi-analytic cases.

Our approach is that of ``modelling a model'', using a semi-analytic model to reproduce the behaviour of the simulation including the effects of limited mass resolution. Since we are interested primarily in the rate at which cooling occurs in the two models, we use a simulation which allows radiative cooling but which does not include any prescription for star formation or feedback. We attempt to model this simulation using a ``stripped down'' semi-analytic model which also neglects these phenomena. Hierarchical models of galaxy formation without feedback predict that most of the gas in the universe cools in low mass objects at high redshift (e.g. \cite{wr78,cole91,wf91}). Consequently, we cannot expect either our SPH simulation or our stripped down semi-analytic model to cool realistic quantities of gas, and where differences between the two approaches are found it may not be possible to conclude that one is more ``correct'' than the other. However, the changes which must be made to the semi-analytic model to match the SPH simulation may provide insight into the level of agreement between the two techniques and the reasons for any discrepancies.

The layout of this paper is as follows. In Section~\ref{sec:themodels} we describe our semi-analytic model and give details of the SPH simulation we use. In Section~\ref{sec:sphvsdsa} we compare properties of the two models, including galaxy masses, cold gas mass in halos as a function of redshift and the spatial distribution of the galaxies. In Section~\ref{sec:conclusions} we present our conclusions.

\section{The Models}
\label{sec:themodels}

\subsection{The SPH Simulation}
\label{sec:sphsim}

SPH is a Lagrangian numerical method which follows the motion of a set of gas elements represented by discrete particles. The thermal energy and velocity of each particle are known at any given time and each particle has a fixed mass. Properties of the gas at the position of a particle can be estimated by smoothing these quantities over the $N_{\rm{SPH}}$ nearest neighbouring particles. The gas properties are then used to calculate the forces acting on each particle in order to update the positions and velocities. In cosmological simulations both dark matter and gas particles are included and the particles are initially distributed in a manner consistent with a cosmological power spectrum. If the process of galaxy formation is to be simulated then radiative cooling of the gas must also be included.

The SPH simulation used here was performed using the Hydra code. This particular implementation includes a modification, described by Pearce \etal (\shortcite{p2001}), to prevent the rate of cooling of hot gas being artificially increased by nearby clumps of cold, dense gas, or ``galaxies''. Any gas hotter than $10^5$K is assumed not to interact with gas at temperatures below $12\,000$K. Thus, for cooling purposes the density estimate for a hot particle near a galaxy is based only on the neighbouring hot particles and the cooling rate is unaffected by the presence of the galaxy.

The simulation has $80^3$ gas and $80^3$ dark matter particles with individual masses of $2.57 \times 10^{9}h^{-1} \rm{M_{\odot}}$ and $2.37 \times 10^{10} h^{-1} \rm{M_{\odot}}$ respectively, contained in a cube of side $50h^{-1}\rm{Mpc}$. The power spectrum is that appropriate to a cold dark matter universe with the following parameter values: mean mass density parameter $\Omega_0=0.35$, cosmological constant $\Lambda_0=0.65$, baryon density parameter $\Omega_b=0.0377$, Hubble constant $h=0.71$, power spectrum shape parameter $\Gamma=0.21$ and rms linear fluctuation amplitude $\sigma_8=0.90$. The gravitational softening length is $25 h^{-1}\rm{kpc}$, fixed in physical coordinates.

The metallicity of the gas in the simulation, measured in terms of the mass fraction of metals, $Z$, is uniform and varies linearly with time according to:
\begin{equation}
Z=0.3 Z_{\rm{\odot}} \, t(z)/t_0,
\label{eqn:metal}
\end{equation}
where $Z_{\rm{\odot}}$ denotes the solar metallicity, $t(z)$ is the age of the universe at redshift $z$ and $t_0$ is the age of the universe at $z=0$.

This simulation makes no attempt to treat star formation and does not include any heating or feedback processes.

\subsection{The N-body GALFORM Model}
\label{sec:model}

The semi-analytic model used here, which we will refer to as N-body \galf, is a galaxy formation model which uses the output from an N-body simulation to calculate halo merger histories and semi-analytic techniques to model baryonic processes. Briefly, halo merger trees are constructed by identifying halos at each simulation output time using the friends-of-friends (FOF) algorithm of Davis \etal (\shortcite{davis85}). Each halo at each output time is identified as a progenitor of whichever halo contains the largest fraction of its mass at the next output time. The merger history of each halo at the final time can then be traced back. Semi-analytic techniques are used to treat the shock heating of gas during the formation of a halo, the cooling of gas within halos and, in the general case, the formation of stars and the merging of galaxies within halos. The full model predicts a wide range of galaxy properties including luminosity, stellar masses of the bulge and disk components and cold gas mass. Galaxy positions can be obtained since each galaxy is associated with a particle in the N-body simulation. Initially, this will be taken to be the most bound particle of the halo in which the galaxy formed, but if the galaxy subsequently merges with the central galaxy of another halo it will be associated with the most bound particle of that halo.

The semi-analytic methods employed in this work are taken from the \galf\ model of Cole \etal (\shortcite{cole2k}), who use a Monte-Carlo algorithm to generate realisations of halo merger histories. Helly \etal (\shortcite{paperI}) describe the N-body \galf\ model and the technique used to obtain merger trees in detail, and investigate the effects of using simulation derived merger trees on the predicted galaxy populations. 

In order to allow a direct comparison between the predictions of this model and those of the SPH simulation, the merger trees must be calculated from the dark matter component of the SPH simulation. Consequently, the time and mass resolution in the halo merger trees are determined by the properties of the SPH simulation and differ from the time and mass resolution of the simulation employed by Helly \etal We have a total of 61 outputs from the SPH simulation, the first 26 of which are logarithmically spaced in expansion factor between redshifts $z \sim 10$ and $z \sim 1.5$. The remaining outputs are equally spaced in time between $z \sim 1.5$ and $z=0$. This is something of an improvement in time resolution over the simulation used by Helly \etal (\shortcite{paperI}). However, the predictions of the \galf\ model were not significantly affected when the number of timesteps was increased, so we do not expect this difference to be important. 
 
There are two parameters which we vary in order to model the SPH simulation. The N-body \galf\ model assumes that the distribution of mass in dark matter halos is described by the radial density profile found by Navarro, Frenk \& White (\shortcite{nfw96}, \shortcite{nfw97}). This profile contains a single free parameter, which can be expressed as the concentration parameter, $c$, defined by Navarro, Frenk \& White or a halo scale radius, $r_{\rm{NFW}}=r_{\rm{virial}}/c$, where $r_{\rm{virial}}$ is the virial radius of the halo. Like Cole \etal (\shortcite{cole2k}), we set $r_{\rm{NFW}}$ using the method described in the appendix of Navarro, Frenk \& White (\shortcite{nfw97}). No scatter is included in the scale radius as a function of halo mass. The radial density profile we assume for the hot gas within halos is described by Helly \etal This profile also contains one parameter, the core radius $r_{\rm{core}}$, which we specify as a fraction of $r_{\rm{NFW}}$ and may be held at a fixed value or allowed to increase with time from an initial value $r_{\rm{core}}^0$. See Helly \etal (\shortcite{paperI}) for details. 

We also allow ourselves the freedom to vary the rate at which mergers occur between galaxies in the same dark matter halo. This is specified in terms of a merger timescale parameter, $f_{\rm{df}}$, which is a prefactor in the standard dynamical friction timescale. Reducing $f_{\rm{df}}$ increases the rate at which mergers occur. See Cole \etal (\shortcite{cole2k}) for details of the merger scheme we use.

\section{Comparison between SPH and N-body GALFORM}
\label{sec:sphvsdsa}

\begin{figure*}
\epsfxsize=18.5 truecm \epsfbox{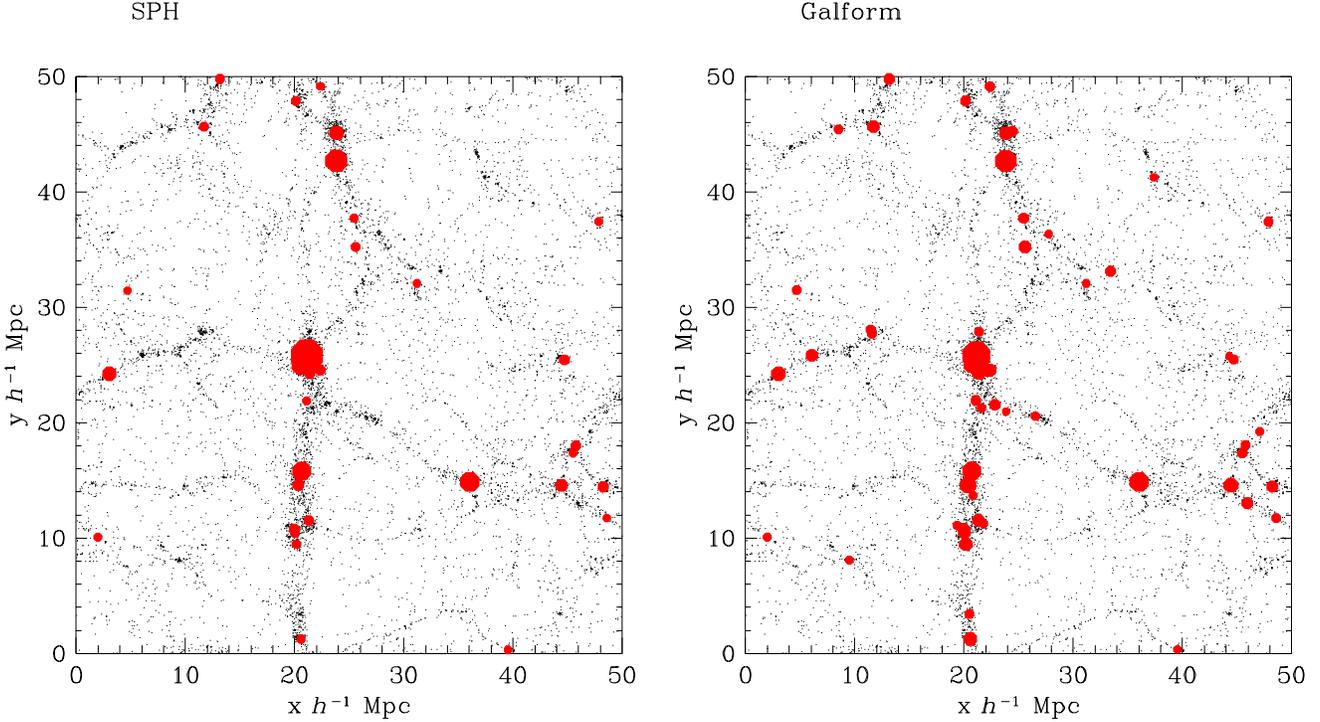}
\caption{Positions and masses of galaxies in a $5h\rm{^{-1}Mpc}$ thick slice through the simulation volume. The panel on the left shows galaxies found in the SPH simulation using a friends of friends algorithm to identify clumps of cold gas particles. The panel on the right shows the galaxies predicted by the N-body \galf\ model. Each circle represents a galaxy, and the area is proportional to the mass of the galaxy. Dark matter particles are shown as dots. Only galaxies with masses greater than 32 gas particle masses, or $8.2\times10^{10}h^{-1}\rm{M_{\odot}}$, are shown. }
\label{fig:dotplot}
\end{figure*}

In this section we compare the results of the SPH simulation with the N-body \galf\ model, which uses merger trees derived from the dark matter component of the SPH simulation. Fig.~\ref{fig:dotplot} shows the positions and masses of the galaxies which form in a $5h\rm{^{-1}Mpc}$ thick region in both the SPH simulation and N-body \galf\ . The SPH ``galaxies'' (i.e. clumps of cold gas) shown here were identified using a FOF group finder on gas particles with temperatures between $8\,000$ and $12\,000$K (see Section~\ref{sec:halobyhalo}). 

\subsection{Modelling SPH with N-body GALFORM}
\label{sec:comparison} 

In order to produce a semi-analytic model of the SPH gas simulation using N-body \galf\, we must first remove the treatment of star formation, feedback and chemical enrichment from \galf. We set the metallicity of the gas to be the same as that in the simulation, using eqn.~(\ref{eqn:metal}).

The cooling rate of the gas in our simulation depends on its density, which is estimated by searching for the $N_{\rm{SPH}}$ nearest neighbours. The density of gas in halos with less than $N_{\rm{SPH}}=32$ gas particles, or a total gas mass less than $8.2\times 10^{10} h^{-1}\rm{M_{\odot}}$, will in general be severely underestimated with an associated suppression of the cooling rate. Consequently, the mass of gas which cools is dependent on the particle mass.

\begin{figure}
\epsfxsize=8.5 truecm \epsfbox{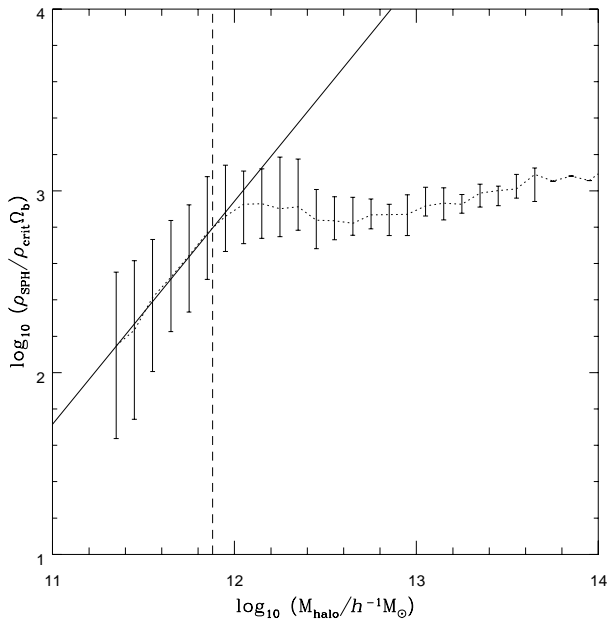}
\caption{Mean halo gas density $\rho_{\rm{SPH}}$ plotted against halo mass $M_{\rm{halo}}$ at redshift $z=0$. The density is expressed in terms of the universal baryon density. The mean density is calculated from density estimates for individual particles in the SPH simulation. The dotted line shows the median of the mean halo gas densities as a function of halo mass. The error bars show 10 and 90 percentile limits. The vertical dashed line is at a halo mass corresponding to 32 dark matter particles. The solid line is a power law fit to the median density for halos of fewer than 32 particles.}
\label{fig:densityfit}
\end{figure}

In order to model this effect in the semi-analytic treatment, we first investigate the variation of the mean estimated density of gas in halos in the SPH simulation with halo mass. A characteristic volume for each gas particle can be obtained by dividing its mass by its SPH density estimate. The total volume of the gas in a halo is calculated by summing the volumes of its constituent gas particles. The total volume is then divided by the mass of gas in the halo to obtain an estimate of the mean gas density. Fig.~\ref{fig:densityfit} shows this density estimate plotted against halo mass, at redshift $z=0$. In halos identified using the FOF group finder with $b=0.2$ we expect the mean gas density to be several hundred times the universal mean gas density. The dotted line shows the median of the mean densities of halos of a given mass. Halos with more than 32 particles have approximately constant mean density, although the density does increase somewhat with halo mass. 

The estimated density rapidly drops once the halo mass falls below 32 dark matter particle masses. Since the cooling time of the gas is inversely proportional to its density this could significantly affect the amount of gas which cools in the smaller halos in the simulation. We incorporate this effect into the semi-analytic model by increasing the cooling time for gas in halos of fewer than 32 particles. A least squares fit to Fig.~\ref{fig:densityfit} gives:
\begin{equation}
\log_{10} \frac{\bar{\rho}_{\rm{SPH}}}{\rho_{\rm{crit}} \Omega_b} = 1.23 \log_{10} \rm{M}_{\rm{halo}} - 11.79,
\label{eqn:densityfit}
\end{equation} where $\bar{\rho}_{\rm{SPH}}$ is the mean gas density estimated from the SPH simulation and $\rm{M}_{\rm{halo}}$ is the mass of the halo. The cooling time in our model is inversely proportional to the mean density of the gas in the halo. In halos of fewer than 32 particles we replace the cooling time, $\tau_{\rm{cool}}$, with a longer cooling time, $\tau_{\rm{cool}}^{\rm{SPH}}$, given by
\begin{equation}
\tau_{\rm{cool}}^{\rm{SPH}} = k \, \tau_{\rm{cool}} \frac{\Omega_b\,\rho_{\rm{crit}}}{\bar{\rho}_{\rm{SPH}}},
\label{eqn:adjustrho}
\end{equation} where $\rho_{\rm{crit}}$ is the critical density. We set the constant of proportionality, $k$, in this relation by requiring that the cooling time for halos of 32 particles be unchanged.

\subsubsection{Halo by halo comparison}
\label{sec:halobyhalo}

The masses of individual galaxies in the N-body \galf\ model depend on the rate at which galaxy mergers occur within dark matter halos. Since the merger rate in the SPH simulation may not be the same as that in the semi-analytic model, we first compare the total amount of gas which cools in halos of a given mass. This quantity should be independent of the merger rate, at least in the semi-analytic case, and can be used to compare the treatment of cooling in the two models. In the SPH simulation a large galaxy forming at the centre of a halo through mergers may gravitationally affect the density, and hence the cooling rate, of nearby gas, but we do not expect this to be a large effect and the mass of gas which cools should be only weakly dependent on the merger rate.

We adopt two different models for the evolution of the gas density profile in the semi-analytic treatment. The first is that used by Cole \etal (\shortcite{cole2k}) in which the core radius in the gas profile increases with time in order to maintain the gas density at the virial radius. We may vary the initial core radius, $r_{\rm{core}}^0$, in order to adjust the amount of gas which cools (the standard choice adopted by Cole \etal was $r_{\rm{core}}^0 = 0.33 r_{\rm{NFW}}$. The second is a simpler model in which the core radius remains a constant fraction of the halo scale radius, $r_{\rm{NFW}}$. Again, the size of this fixed core may be varied in order to adjust the rate at which cooling occurs.

In order to quantify the mass of cold gas present in halos in the SPH simulation, we first associate gas particles with dark matter halos. A gas particle is considered to belong to a halo if it lies within a linking length $b=0.2$ of a dark matter particle which belongs to that halo. In the unlikely event that dark matter particles from more than one halo are found within the linking length, the gas particle is assigned to the halo containing the nearest dark matter particle. The linking length used in this procedure is the same as that used to identify dark matter halos with the FOF group finder. This ensures that the condition for a gas particle to be associated with a halo is consistent with the definition of halo membership used for the dark matter particles.

The cooling function in our simulation permits gas to cool only to a temperature of $10^{4}$K. This allows us to distinguish between gas which has been heated and has subsequently cooled to $10^{4}$K and the diffuse cold gas in voids which has never been heated and is at much lower temperatures. The mass of gas which has cooled in each halo is obtained by summing the masses of all gas particles associated with the halo and having temperatures between $8\,000$K and $12\,000$K. In the N-body \galf\ model, the amount of cold gas in each halo is simply the mass of gas which has cooled from the hot phase, since the model includes no star formation.

\begin{figure*}
\epsfxsize=18.5 truecm \epsfbox{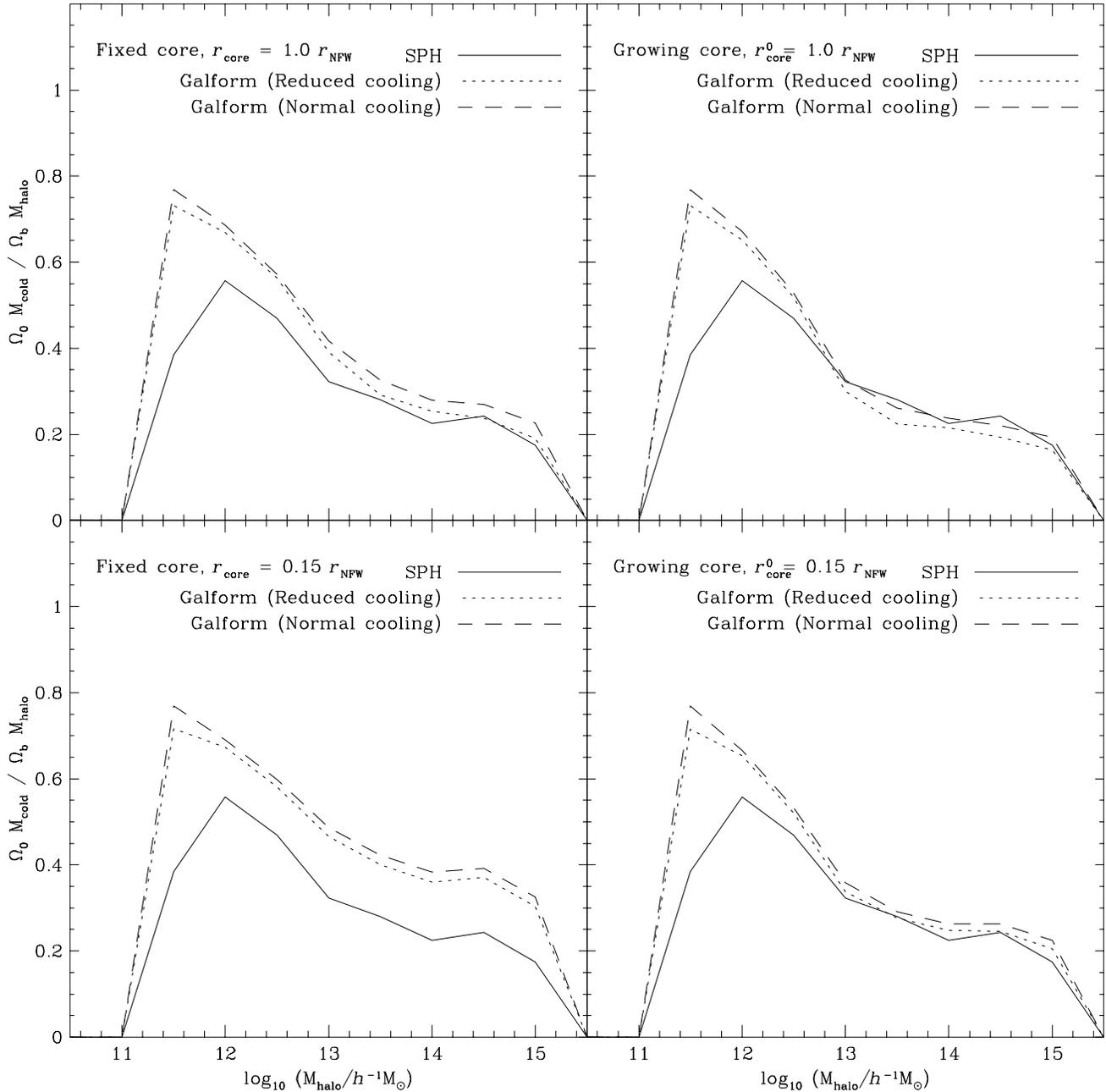}
\caption{Mean fraction of halo gas which has cooled at redshift $z=0$ as a function of halo mass. The solid lines show the mean cooled gas fraction in halos in the SPH simulation and are the same in all four panels. The dotted lines show the cold gas fraction in N-body \galf\ models where the cooling time in low mass halos is increased according to eqn~(\ref{eqn:adjustrho}). The dashed lines show N-body \galf\ models without this adjustment. In the upper panels the initial core radius is set equal to the NFW scale radius of the halo. In the lower panels the core radius is set to 0.15 times the scale radius. In the panels on the left hand side the core radius remains fixed at its initial value for all redshifts, in the panels on the right it is allowed to increase to maintain the density of gas at the virial radius.} 
\label{fig:coldfrac}
\end{figure*}

Fig.~\ref{fig:coldfrac} shows the mean fraction of gas which has cooled as a function of halo mass, in both N-body \galf\ and the SPH simulation. Here we consider four different N-body \galf\ models. We vary the initial core radius in the gas profile between $r_{\rm{core}}^0=1.0 r_{\rm{NFW}}$ and $0.15 r_{\rm{NFW}}$ and either fix the core radius as a fraction of the NFW scale radius or allow it to increase with time as described earlier. In the case of a fixed core, $r_{\rm{core}}=r_{\rm{core}}^0$ at all times.  

The dotted lines in Fig.~\ref{fig:coldfrac} show N-body \galf\ models which include the modification to the cooling time in low mass halos described by eqn.~(\ref{eqn:adjustrho}). All four models reproduce the quantities of cold gas observed at redshift $z=0$ in the SPH simulation remarkably well, for halos of mass greater than about $10^{12}h^{-1}\rm{M_{\odot}}$ or around 40 dark matter particles --- in all but the worst case the difference is less than 50\%. We find that if the core radius in the gas density profile is allowed to increase as gas cools, the fraction of cold gas is not particularly sensitive to the choice of intitial core radius, although a small initial value, $r_{\rm{core}}^0=0.15 r_{\rm{NFW}}$, gives a slightly better match than if the core is initially larger. If the core radius is fixed as a fraction of the NFW scale radius a much larger value, $r_{\rm{core}}^0=1.0 r_{\rm{NFW}}$, is necessary.

The dashed lines in the figure show the fraction of gas which cools if cooling is allowed to occur at the normal rate in halos of all masses down to the mass of the smallest halo we can resolve in the simulation. Surprisingly, this appears to have little effect on halos with fewer than 32 dark matter particles for which the cooling rate has been altered. The fraction of gas which has cooled in larger halos also increases by a similar amount. The extra cold gas in these halos must have cooled in progenitors of fewer than 32 particles before being incorporated into larger halos. Overall, the change is not large, with some halos having around 10-20\% more cold gas on average. This suggests that our results are not particularly sensitive to the way in which we model the loss of cooling efficiency in low mass halos, although in both cases the agreement between the SPH simulation and the semi-analytic model is poor in such halos.

\begin{figure*}
\epsfxsize=18.0 truecm \epsfbox{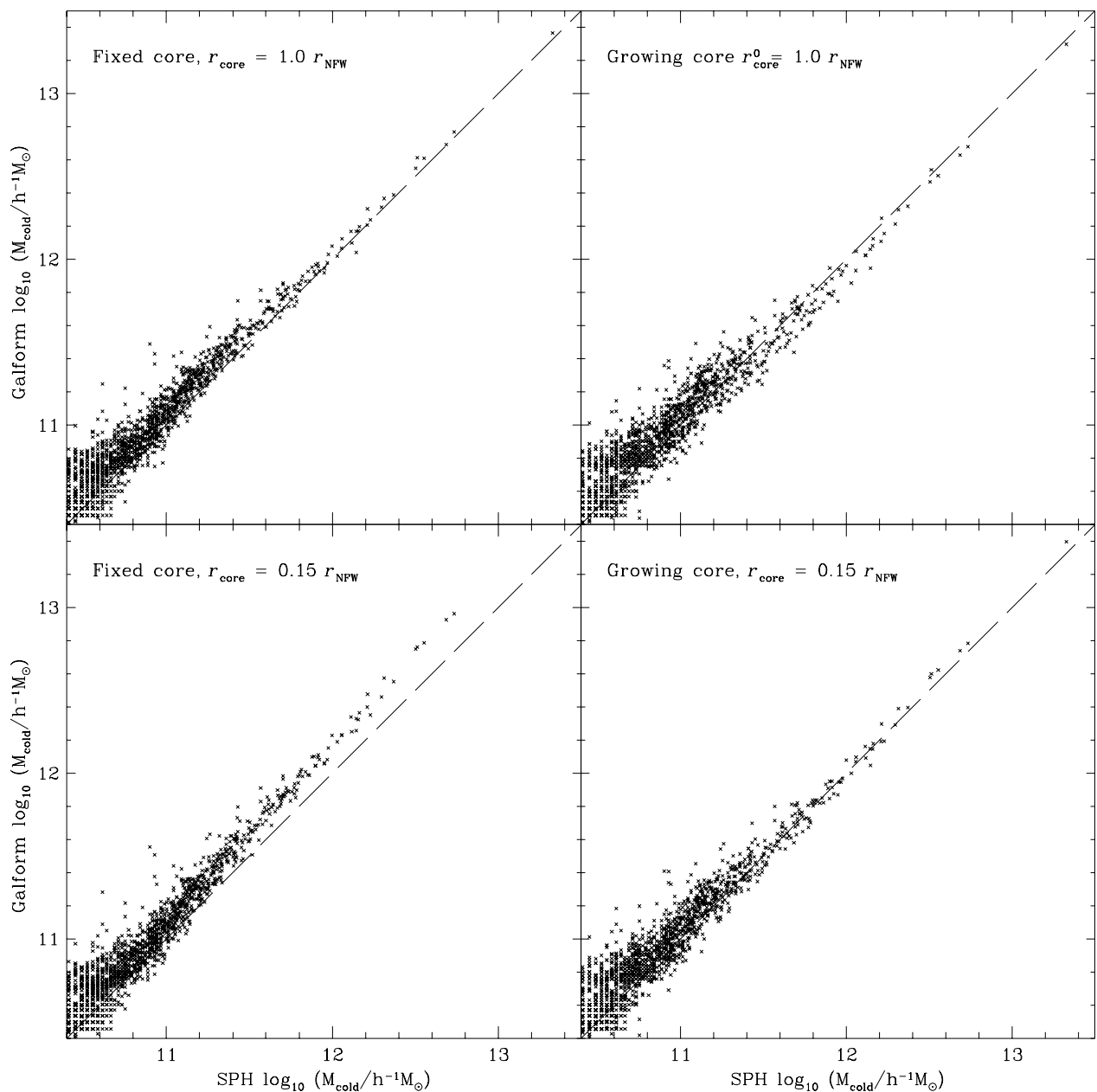}
\caption{Halo cold gas mass, $ M_{\rm{cold}}$, in four different N-body \galf\ models plotted against halo cold gas mass in the SPH simulation at redshift $z=0$. Each point corresponds to a single dark matter halo. The upper panels show N-body \galf\ models with $r_{\rm{core}}^0=1.0r_{\rm{NFW}}$. The lower panels have $r_{\rm{core}}^0=0.15r_{\rm{NFW}}$. In the panels on the left, the core radius in the gas density profile is a fixed fraction of the NFW scale radius. In the panels on the right the core radius is allowed to grow in order to maintain the gas density at the virial radius.}
\label{fig:mcoldsummed}
\end{figure*}

Fig.~\ref{fig:mcoldsummed} shows a direct comparison between the masses of cold gas in individual halos in the SPH simulation and the four N-body \galf\ models of Fig.~\ref{fig:coldfrac}, again using the modified cooling time for low mass halos. The mass of cold gas predicted by N-body \galf\ is plotted against the mass of cold gas in the simulation for each halo, with the initial core radius set to $r_{\rm{NFW}}$ in the upper panels and $0.15r_{\rm{NFW}}$ in the lower panels. In the models shown on the left-hand side the core radius remains fixed at its initial value at all times. The long-dashed lines show where the points would lie if the simulation and the semi-analytic models were in perfect agreement. 

Again, in all four cases the mass of cold gas in the SPH simulation is well correlated with the mass of cold gas in the N-body \galf\ model.  The small scatter, at least at high masses, shows that the dependence of cold gas mass on merger history must be similar in the SPH simulation and the semi-analytic model. N-body \galf\ with a gas density profile with a fixed core radius appears to cool on average more gas in halos of all masses than the SPH simulation. This can be alleviated to some extent by increasing the size of the core in the gas profile but it appears that a rather large core in the gas distribution would be required to obtain good agreement. Allowing the core radius to increase as gas cools reduces the rate of cooling and results in closer agreement with the simulation; the best agreement is obtained for a small initial core radius of around $0.15r_{\rm{NFW}}$, although the mass of cold gas in each halo is clearly not particularly sensitive to the initial core radius in this \galf\ model.  

\begin{figure*}
\epsfxsize=18.5 truecm \epsfbox{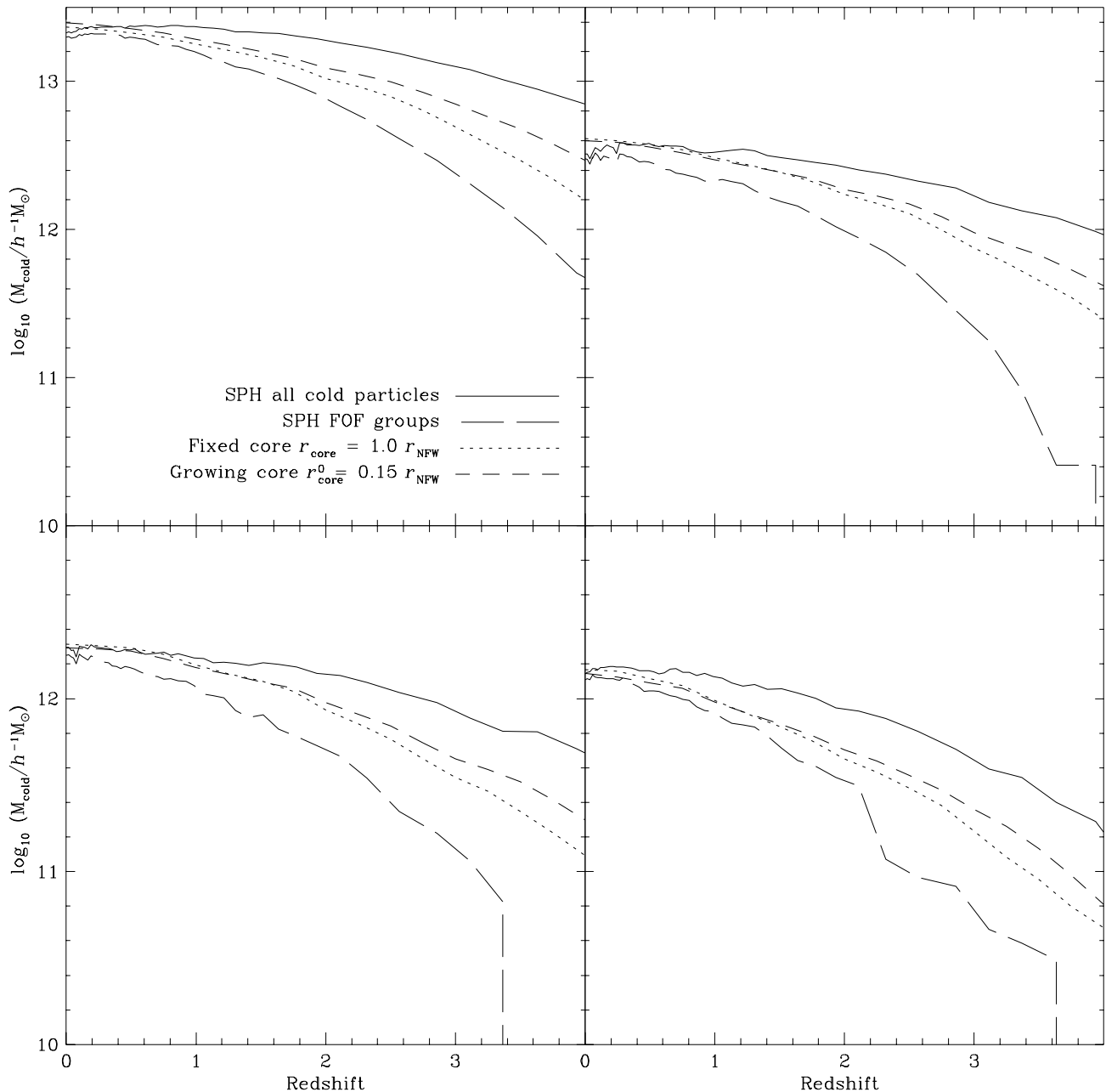}
\caption{Mass of cold gas in the progenitors of four halos as a function of redshift. Each panel corresponds to a single halo at $z=0$. The solid line shows the mass of cold gas in the SPH simulation obtained by summing the masses of all cold gas particles in the progenitors. The long dashed line shows the mass of cold gas obtained by summing the masses of all FOF groups of cold particles in the progenitors. The dotted lines correspond to an N-body \galf\ model with a fixed core radius in the gas density profile with $r_{\rm{core}}=r_{\rm{NFW}}$. The short dashed lines correspond to a model with a growing core radius of initial value $r_{\rm{core}}^0=0.15r_{\rm{NFW}}$.}
\label{fig:mcold_z}
\end{figure*}

Fig.~\ref{fig:mcold_z} shows the mass of cold gas in progenitors of four of the larger halos in the simulation as a function of redshift. The mass of cold gas in the simulation (solid lines) at a given redshift is obtained by summing the masses of all cold gas particles associated with the progenitors of the final halo at that redshift. Particles are associated with halos using the method described earlier in this section and, as before, ``cold'' particles are those with temperatures in the range $8\,000$--$12\,000$K. Similarly, the mass of cold gas in the N-body \galf\ model is obtained by summing the masses of the galaxies in the progenitor halos. Here we show results for two models, one with $r_{\rm{core}}$ fixed at $r_{\rm{core}}^0=1.0r_{\rm{NFW}}$ (dotted lines) and the other with a growing core which has an initial core radius $r_{\rm{core}}^0=0.15r_{\rm{NFW}}$ (dashed lines). The model of Cole \etal used a gas profile with a larger initial core radius, $r_{\rm{core}}^0=0.33r_{\rm{NFW}}$.

The long dashed lines show the mass of cold gas in progenitors in the simulation if instead of associating gas particles with halos directly, we use the FOF group finder to first identify clumps of cold gas and then associate clumps with dark matter halos. A clump is assigned to a halo if a dark matter particle from that halo is found within a dark matter linking length of the clump's centre of mass. If particles belonging to several halos are found in this region, the nearest to the centre of mass is used. A linking length $b=0.02$ is used to identify the clumps and a minimum group size of 10 particles is imposed on the clumps. These lines are shown in Fig.~\ref{fig:mcold_z} only to illustrate that there is some dependence on the way in which we define ``cold halo gas'' in the simulation. This second method will certainly underestimate the mass of cold gas because the group finder imposes a minimum mass on the clumps, missing smaller groups of cold particles. Also, at high redshift the gravitational softening length exceeds the linking length used to identify the clumps, so particles which ought to be considered part of a clump may not have collapsed to sufficiently high densities to be picked up by the group finder. We find that most of the discrepancy between these two SPH results is due to cold particles in small groups of fewer than five particles, at least with $b=0.02$.

It is also possible that the first method of counting individual gas particles associated with halos overestimates the mass of cold gas in smaller halos, where the linking length becomes a significant fraction of the radius of the halo. Any particle within a linking length of the outer dark matter particles of the halo may be associated with that halo. Despite this uncertainty, it appears that more of the cold gas found in the simulation cooled at high redshift than in either of the N-body \galf\ cases considered. At redshift 2 the discrepancy is approximately a factor of 2. Allowing the core radius to increase from a small initial value helps somewhat by encouraging more cooling initially and slightly suppressing it later, but the improvement is small compared to the size of the discrepancy with the SPH simulation for redshifts greater than around 2. Reducing the initial core radius in this model further has little effect on these results.

We have tried to model the effect of limited resolution on cooling in SPH blobs of fewer than 32 dark matter particles, but in the N-body \galf\ model no cooling is possible in halos of fewer than 10 dark matter particles. It appears that in our SPH simulation some cooling \emph{does} occur in these halos. However, it may not be useful to model the rate of cooling in this regime, since it is entirely artificial and likely to be dependent on the details of the particular SPH implementation. In any case, when halos in the SPH simulation first grow to 10 dark matter particles they may have already cooled some gas. These halos will eventually be incorporated into larger halos, where the cold gas mass becomes dominated by material which cooled in well resolved halos so that at late times the SPH and \galf\ calculations converge. 

\subsubsection{Galaxy by galaxy comparison}
\label{sec:galbygal} 

\begin{figure}
\epsfxsize=8.5 truecm \epsfbox{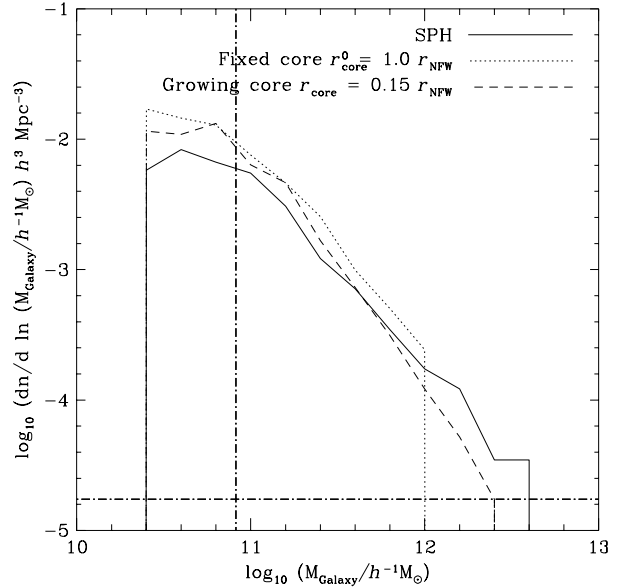}
\caption{Galaxy number density as a function of cold gas mass at redshift $z=0$. The solid line shows galaxy number density in the SPH simulation. The other lines correspond to N-body \galf\ models with 1) a fixed core radius $r_{\rm{core}}=r_{\rm{NFW}}$ (dotted line) and 2) a growing core which initially has $r_{\rm{core}}^0=0.15r_{\rm{NFW}}$ (dashed line). The horizontal dot-dashed line shows the number density equal to one object per simulation volume. The vertical dot-dashed line is at a mass equal to 32 gas particle masses.}
\label{fig:galmassfn}
\end{figure}

Fig.~\ref{fig:galmassfn} shows the number density of galaxies as a function of mass in the SPH simulation and in the N-body \galf\ model at redshift $z=0$. Here, SPH ``galaxies'' are groups of particles identified by the FOF group finder applied to all particles with temperatures in the range $8\,000$--$12\,000$K. We use a linking length $b=0.02$ and impose a minimum group size of 10 particles. The results are insensitive to the specific choice of $b$ within reasonable bounds. Two N-body \galf\ cases are shown, one with a core of fixed size $r_{\rm{core}}=r_{\rm{NFW}}$ in the gas density profile, the other with a growing core of initial size $r_{\rm{core}}^0=0.15r_{\rm{NFW}}$. In both cases N-body \galf\ predicts about 50\% more galaxies with masses around $3 \times 10^{11}h^{-1}\rm{M}_{\odot}$ or less and fewer galaxies with masses greater than this for the latter choice of $r_{\rm{core}}^0$. The deficit in the number of massive galaxies is more apparent in the model with a large, fixed gas core radius. Since we know that the total amount of gas cooled in the semi-analytic models in each halo is similar to the amount that cooled in the simulation (see Fig.~\ref{fig:mcoldsummed}), this suggests that there is more merging occurring in the simulation. This does not necessarily indicate a failure of the semi-analytic model, however, since it is possible that numerical effects in the simulation contribute significantly to the merger rate.

\begin{figure}
\epsfxsize=8.5 truecm \epsfbox{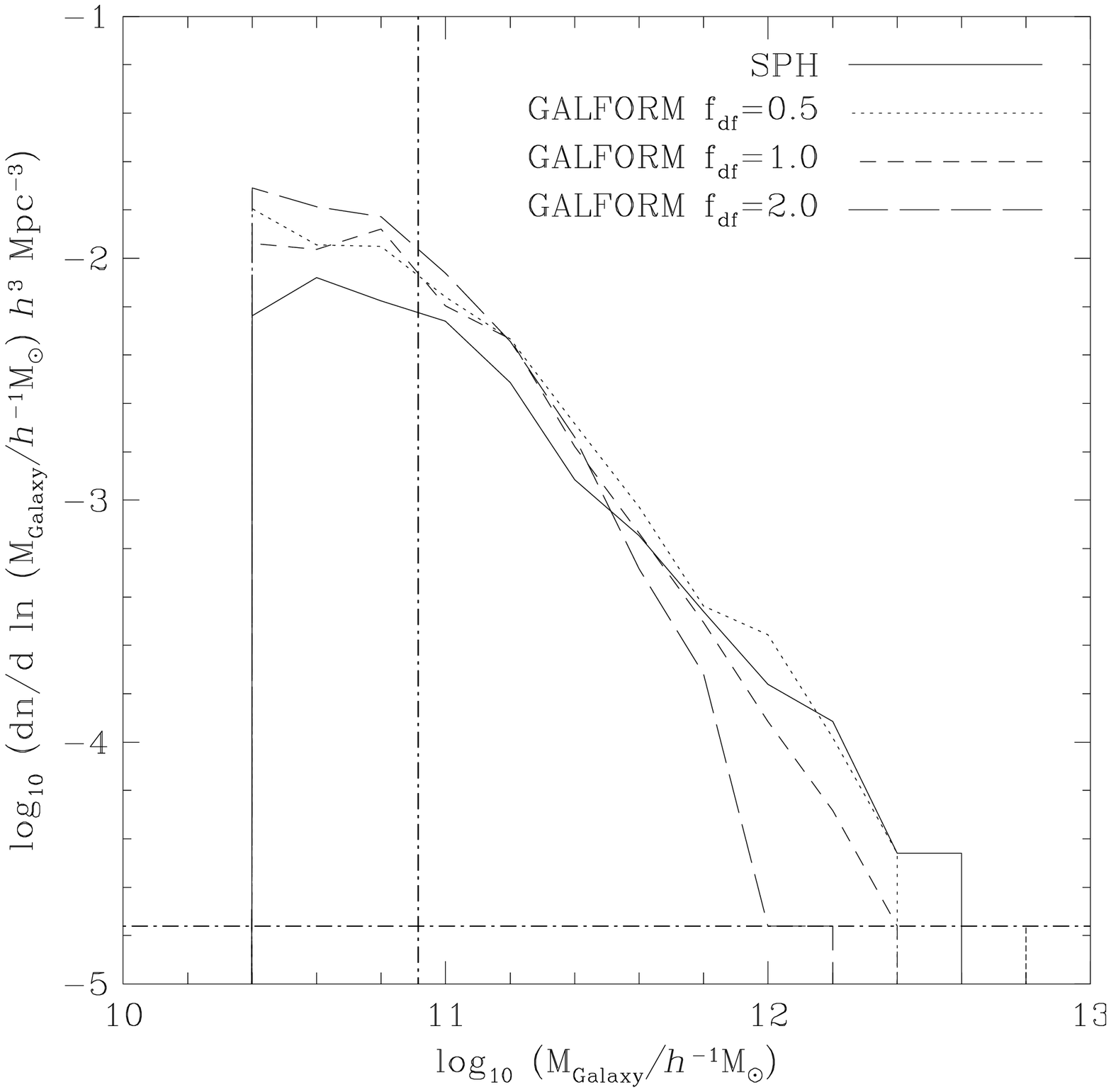}
\caption{Galaxy number density as a function of cold gas mass at redshift $z=0$ for N-body \galf\ models with three different merger rates. All three models have gas profiles with a growing core radius which is initially set to $r_{\rm{core}}^0=0.15r_{\rm{NFW}}$. The merger timescale parameter $f_{\rm{df}}$ is varied between 0.5 (dotted line), 1.0 (short dashed line) and 2.0 (long dashed line). The short dashed line is identical to the short dashed line in Fig.~\ref{fig:galmassfn}. The solid line shows the galaxy number density in the SPH simulation and is identical to the solid line in Fig.~\ref{fig:galmassfn}.  The horizontal dot-dashed line shows the number density corresponding to one object per simulation volume. The vertical dot-dashed line is at a mass equal to 32 gas particle masses. The curves are truncated at 10 gas particle masses.}
\label{fig:vary_tau}
\end{figure}

To test this hypothesis, we varied the merger timescale parameter, $f_{\rm{df}}$ in the semi-analytic models. Fig.~\ref{fig:vary_tau} shows galaxy number density as a function of mass for three N-body \galf\ models with $f_{\rm{df}}$ = 0.5, 1.0 and 2.0. All three have gas profiles with growing cores of initial radius $r_{\rm{core}}^0=0.15r_{\rm{NFW}}$. Doubling the merger timescale ($f_{\rm{df}}=2.0$) drastically reduces the number of more massive galaxies and prevents the formation of any galaxies more massive than $10^{12}h^{-1}\rm{M}_{\odot}$. Halving the merger timescale ($f_{\rm{df}}=0.5$) improves agreement with the simulation by increasing the masses of the largest galaxies and reducing the number of small galaxies. However, the improvement is relatively small and, in any case, the treatment of mergers in the N-body \galf\ model reproduces the distribution of masses observed in the simulation reasonably well with our default $f_{\rm{df}}=1.0$.

The N-body \galf\ model described in Section~\ref{sec:sphvsdsa} does not allow semi-analytic galaxies to be compared with their SPH counterparts on a one to one basis because mergers between galaxies in N-body \galf\ are treated in a statistical manner. While the agreement between the galaxy mass distributions suggests that the overall merger rate in the N-body \galf\ model is similar to that seen in the simulation, we cannot expect mergers to occur between the same galaxies in the two cases, and hence it is not possible to identify clumps of cold gas particles with individual semi-analytic galaxies. 

This problem could be avoided by following the substructure within dark matter halos to determine when mergers between galaxies occur, using a method similar to that of Springel, White, Tormen \& Kauffmann (\shortcite{swtk2001}). Unfortunately the halos in our simulation typically contain too few particles for this to be practical. Any dark matter substructure is rapidly destroyed by numerical effects. 

In order to compare the masses of individual galaxies directly, we need an alternative way to ensure that the same galaxies merge in each model. We do this by using information from the baryonic component of the SPH simulation to merge N-body \galf\ galaxies. We first populate the simulation volume with galaxies calculated using the N-body \galf\ model, with merging of galaxies completely suppressed. We find the halo in which each semi-analytic galaxy first formed, and identify the gas particles associated with that halo as those with indices corresponding to the indices of the dark matter particles in the halo --- this is possible because in our SPH simulation gas and dark matter particles with the same indicies are initially at the same locations and tend to remain in the same halos at later times. By redshift $z=0$ some of these particles will be contained within SPH galaxies. Each semi-analytic galaxy is assigned to the SPH galaxy which contains the largest number of gas particles from the halo in which it formed. This procedure often results in several semi-analytic galaxies being assigned to the same blob of cold gas at redshift $z=0$. These galaxies are assumed to have merged and their masses are added together. It is possible to think of rare situations where our method might incorrectly merge galaxies, but this is the best that can be done within the limitations of the SPH simulation.

We are only able to detect SPH galaxies with 10 particles or more, so it is inevitable that sometimes a semi-analytic galaxy will not be assigned to any SPH galaxy. This would occur if the semi-analytic galaxy formed in a halo which, in the simulation, failed to cool enough particles to constitute a group by redshift $z=0$. Such galaxies are generally found in small, recently formed halos and typically have masses of around 10 gas particle masses or less. These galaxies account for about 20\% of the total semi-analytic galactic mass in the simulation volume. We also find that a small number (about 2\%) of the SPH galaxies have no corresponding semi-analytic galaxy. Almost all of these are poorly resolved objects close to the 10 particle threshold.

\begin{figure*}
\epsfxsize=18 truecm \epsfbox{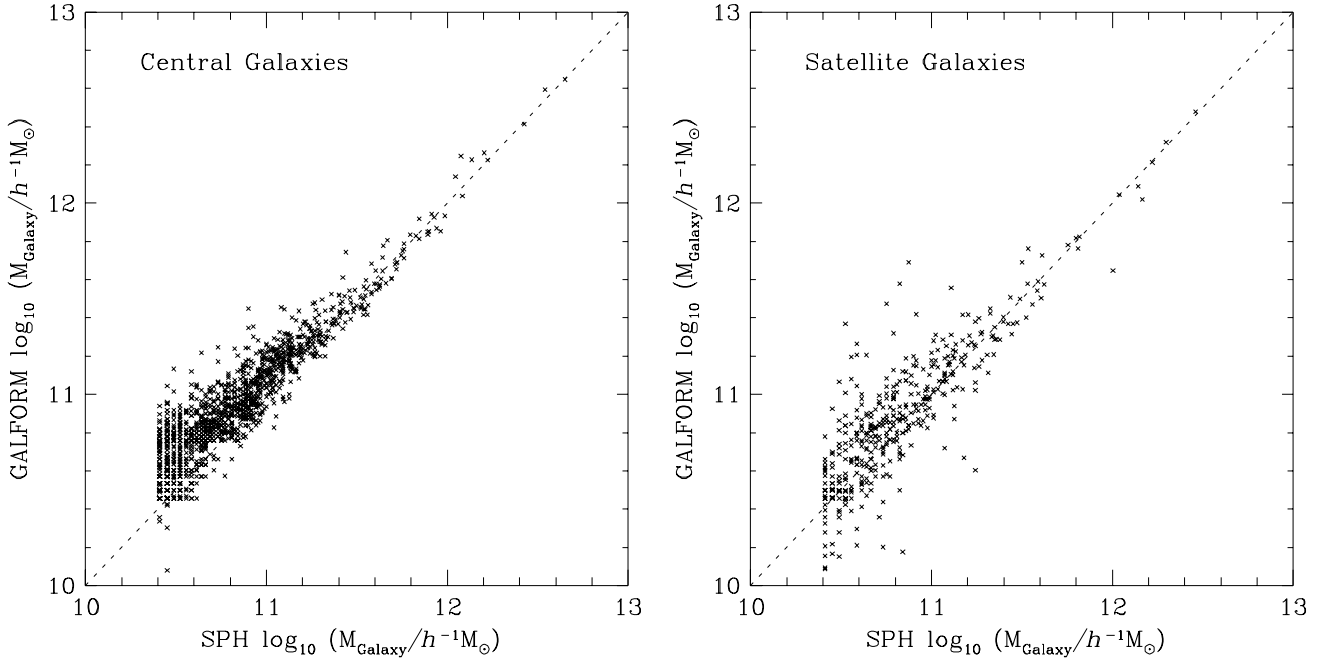}
\caption{Comparison between galaxy masses in the SPH and N-body \galf\ models. The merger scheme described in Section~\ref{sec:galbygal} is used to identify N-body \galf\ galaxies with SPH galaxies. Galaxies lying on the dashed line have equal masses in both models. The panel on the left shows only galaxies which are considered to be central galaxies in the N-body \galf\ model. The panel on the right shows only galaxies which are satellites in the N-body \galf\ model. }
\label{fig:masscomparison}
\end{figure*}

Since the unmatched semi-analytic galaxies largely correspond to SPH galaxies which have yet to gain enough cold particles to be identified by the group finder, we simply omit them from the comparison shown in Fig.~\ref{fig:masscomparison}. Here, we compare the masses of the merged semi-analytic galaxies with the corresponding galaxies in the SPH simulation. Each point on the plot represents a single SPH galaxy which has been associated with one or more semi-analytic galaxies. We have split the galaxies into two categories -- central galaxies (left panel) and satellite galaxies (right panel). This allows us to test the assumption made in the \galf\ model that no gas cools onto satellite galaxies. If this is not true, galaxies which are considered to be satellites in the N-body \galf\ model will have systematically lower masses than their SPH counterparts. It therefore makes sense, for this purpose, to use information from the semi-analytic model (and not the SPH simulation) to determine whether each galaxy is a satellite. The semi-analytic mass of each galaxy shown in Fig.~\ref{fig:masscomparison} is the sum of the masses of the \galf\ galaxy fragments which have been associated with the corresponding SPH galaxy. We identify the galaxy as a central galaxy if any one of those fragements was a central galaxy before we applied our SPH merging scheme. If all of the fragments were satellites, the galaxy is considered to be a satellite.

There is clearly a very strong correlation between the mass of each simulated galaxy and its semi-analytic counterpart, although the N-body \galf\ galaxies appear to be systematically more massive by up to 25\% at low masses. The scatter in this plot is comparable to that in Fig.~\ref{fig:mcoldsummed}. There appears to be little or no systematic difference between satellite and central galaxies, which suggests that no significant amount of cooling of gas onto satellite galaxies is occurring in the simulation. There are a few outlying points where the \galf\ and SPH masses are drastically different -- these are mainly satellites, but there are as many with much \emph{higher} \galf\ masses than SPH masses as there are with lower masses. These are most likely a result of the SPH merging algorithm assigning \galf\ galaxy fragments to the wrong SPH galaxy.


\begin{figure*}
\epsfxsize=18.0 truecm \epsfbox{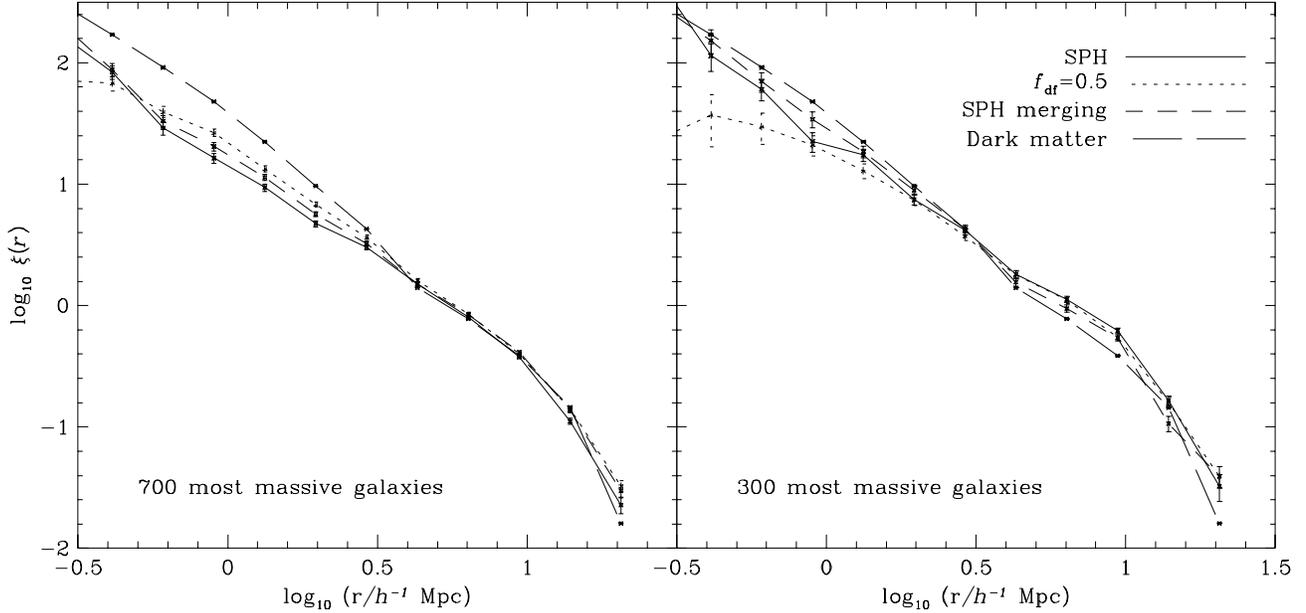}
\caption{Two point galaxy correlation functions for three different models - the SPH simulation (solid lines), an N-body \galf\ model with merger rate parameter $f_{\rm{df}}=0.5$ (dotted lines) and an N-body \galf\ model using the SPH based merger scheme described in Section~\ref{sec:galbygal} (short dashed lines). The long dashed lines show the correlation function for the dark matter in the SPH simulation. The 700 most massive galaxies in each case are included in the calculation for the left panel and only the 300 most massive galaxies are included in the right panel. Both N-body \galf\ models have a gas density profile with a core radius which is allowed to grow from an initial value of $r_{\rm{core}}^0=0.15r_{\rm{NFW}}$.}
\label{fig:correlation}
\end{figure*}

Finally, we compare the clustering of galaxies in the two models. While the spatial distribution of dark matter halos in the N-body \galf\ model is identical to that in the simulation, the number of galaxies in each halo and their distribution within the halo may differ. Fig.~\ref{fig:correlation} shows two point galaxy correlation functions for galaxies in the SPH simulation and two different N-body \galf\ models, both of which have gas profiles with growing core radii which are initially set to $r_{\rm{core}}^0=0.15r_{\rm{NFW}}$. In the first \galf\ model, merging between galaxies is treated using the dynamical friction approach of Cole \etal with $f_{\rm{df}}=0.5$, which gives a closer match to the distribution of galaxy masses in the simulation than our default value of 1.0 (see Fig.~\ref{fig:vary_tau}.) In the second \galf\ model, we use the SPH based merging scheme described earlier in this section and put each merged \galf\ galaxy at the position of its associated SPH galaxy. In each case, we include only the 700 (left panel of Fig.~\ref{fig:correlation}) or 300 (right panel) most massive galaxies in our calculation. This ensures that the overall density of galaxies in the volume is the same in each sample. Picking the 700 largest galaxies excludes all galaxies less massive than about $8\times10^{10}h^{-1}\rm{M_{\odot}}$ or 30 gas particles. Picking the 300 largest galaxies corresponds to a minimum mass of approximately $1.5\times10^{11}h^{-1}\rm{M_{\odot}}$ or around 60 gas particles.

The correlation function has been calculated on scales of up to 25$h^{-1}$Mpc. This is half of the size of the simulation box, so the results presented here should not be treated as predictions of the true galaxy correlation function. Instead, the plots in Fig.~\ref{fig:correlation} are intended to compare the relative clustering of \galf\ and SPH galaxies in our \emph{small} simulation volume. All three models show qualitatively similar behaviour. When we consider the larger sample of galaxies (left panel in Fig.~\ref{fig:correlation}), we see an anti-bias relative to the dark matter on scales of less than a few $h^{-1}$Mpc, with galaxies tracing the dark matter on larger scales. This behaviour agrees with previous semi-analytic (e.g. \cite{kauffmann99,b2000}) and SPH simulation (e.g. \cite{p2001}) results. If we include only the 300 most massive galaxies in the simulation volume (right panel in Fig.~\ref{fig:correlation}), we see that on large scales these more massive galaxies are more strongly clustered than the dark matter in all three cases. 

The N-body \galf\ model with $f_{\rm{df}}=0.5$ is in close agreement with the SPH simulation on scales larger than a few $h^{-1}\rm{Mpc}$ when we use the 700 most massive galaxies. This is to be expected since we have the same distribution of dark matter halos in each case and the merger rate in the semi-analytic model has been adjusted to reproduce roughly the distribution of galaxy masses in the simulation. On length scales smaller than this, where the correlation function is sensitive to the details of our treatment of galaxy mergers within halos, there is a difference of almost a factor of 2 between the SPH simulation and the \galf\ model with  $f_{\rm{df}}=0.5$. The treatment of mergers in this model reproduces the overall distribution of galaxy masses but the merger rates and galaxy distributions in halos of a given mass may not be in close agreement. When we merge \galf\ galaxies by associating them with groups of cold gas in the SPH simulation (short dashed lines in Fig.~\ref{fig:correlation}), the correlation functions agree to within about 25\% on these small scales. If we consider only the 300 most massive galaxies in each case, the correlation function for the model with $f_{\rm{df}}=0.5$ drops to almost an order of magnitude below that of the SPH simulation on scales of about $0.3h^{-1}\rm{Mpc}$. Again, this is due to differences in the merger rates in halos of a given mass since the discrepancy disappears if we use our SPH-based merging algorithm.  

Once we ensure that the same galaxies merge in each model, any remaining differences between the correlation functions shown must be due to differences in the galaxy masses. The most massive 700 objects in the SPH model must be a somewhat different population to the 700 most massive objects in the semi-analytic model. In fact, we find that the two samples possess only 590 objects (about 85\%) in common. This is an inevitable consequence of the scatter in the relation between SPH and semi-analytic galaxy masses shown in Fig.~\ref{fig:masscomparison}. Unless there is zero scatter, there will always be galaxies just massive enough to be included in the correlation function for one model which will not be included in the sample for the other. This explains why the level of agreement is reduced when we consider only the most massive galaxies, where we might have expected to obtain improved agreement. By increasing the minimum mass required for a galaxy to be included in each sample we increase the proportion of galaxies which have masses close to the threshold and the fraction of galaxies common to both samples falls slightly to 237 out of 300, or about 80\%.

\section{Discussion and Conclusions}

In this paper we have used the N-body \galf\ model of Helly \etal (\shortcite{paperI}) to compare the results of a semi-analytic calculation of the radiative cooling of gas in halos with results from a cosmological SPH simulation. We have tried to reproduce the results of the simulation by adjusting the semi-analytic cooling prescription and modelling the effects of limited mass resolution on the SPH cooling rate.

We compared properties of halos in the simulation with the properties of the same halos in the N-body \galf\ model. First, we looked at a global property of the halo population, the average fraction of cooled gas at redshift $z=0$ as a function of halo mass. We found that a model in which the gas density profile with an initially small core radius which is able to increase with time provided the best match to the mean cold gas fractions seen in the SPH simulation among those considered. The level of agreement was excellent for halos with masses above the resolution limit of the SPH simulation.

Our method also enabled us to compare the cool gas content of individual halos. For the gas density profile described above, and also for a profile with a fixed core radius, the total mass of cold gas in each halo was found to be in remarkably good agreement at cold gas masses greater than about $10^{12}h^{-1}\rm{M_{\odot}}$. In poorly resolved halos with lower cold gas masses the scatter in this comparison increased substantially, to a factor of about 3. We found that much of the cold gas found in the more massive halos in the N-body \galf\ model generally cooled at later times than the gas in the same halos in the SPH simulation. By a redshift of 2 in the N-body \galf\ case, the progenitors of the halos contained only half as much cold gas as was present in the simulation. As the redshift increases, the mass of cold gas in the SPH simulation becomes dominated by material which cooled in very small halos, where the cooling rate may be strongly affected by resolution effects and depends sensitively on the SPH implementation (\cite{s2002}). These effects are difficult to model reliably and so the discrepancy between the \galf\ and SPH cold gas masses increases at higher redshifts.

We then turned our attention to the properties of individual ``galaxies'' (i.e. cold gas clumps) at redshift $z=0$. Our best fit model gave a distribution of galaxy masses in good agreement with those in the SPH simulation for galaxies of more than 32 particles when we used the merger timescale of Cole \etal (\shortcite{cole2k}), although the N-body \galf\ model contained a somewhat greater number of low mass galaxies and fewer very massive galaxies than the simulation. Doubling the merger rate in the \galf\ model improved the agreement at all masses, but note that the merger rates in the SPH simulation may not be reliable due to the effects of artificial viscosity (\cite{frenk96}).

In our semi-analytic approach, galaxy mergers are treated in a probabilistic fashion based on the dynamical friction timescale. Thus, a direct identification of semi-analytic and SPH galaxies is not possible. In order to circumvent this problem, we suppressed all merging in the N-body \galf\ model and then used information from the SPH simulation to merge the semi-analytic galaxies and to associate the merged galaxies with groups of cold gas particles in the simulation. This gave us a semi-analytic mass for each galaxy in the SPH simulation. We found that these masses were generally similar (within about 50\% for larger galaxies) with a scatter close to that seen in the comparison of halo cold gas masses.

Finally, we examined the clustering properties of the more massive galaxies in the SPH simulation and two N-body \galf\ models. The first used the dynamical friction treatment of galaxy mergers, the second used our SPH merging scheme. We found that the correlation functions of galaxies in both \galf\ models agreed well with the SPH simulation on scales larger than typical group and cluster sizes, but that on scales of a few $h^{-1}$Mpc or less the correlation function of galaxies in the \galf\ model with merging based on the dynamical friction timescale was higher by almost a factor of 2. Using the SPH merging scheme reduced this discrepancy to about 25\%. 

Our comparison shows that it is possible to reproduce accurately gas cooling, and to a lesser extent galaxy merger rates, in an SPH simulation using semi-analytic methods. Benson \etal (\shortcite{b2001}) demonstrated that the overall rate of cooling, globally and in halos of a given mass, predicted by SPH and semi-analytic models show remarkable consistency. They found that the overall fractions of hot gas, cold, dense gas and uncollapsed gas agreed to within 25\% at $z=0$. The cold gas fractions in halos of a given mass were found to agree to within 50\%, with the SPH simulation cooling more gas than the semi-analytic model. This is consistent with the results presented here, since our best semi-analytic model assumes a gas density profile with a smaller core radius than that of Benson \etal, resulting in a higher central gas density in each halo and more rapid cooling.

 Here we have shown that, with only minor changes to the semi-analytic model, very close agreement can be obtained on a halo by halo basis when merger trees are taken from the SPH simulation. The agreement between SPH and semi-analytic masses for individual halos indicates that the dependence of the cooling rate on merger history is very similar in the two cases. Given the quite different limitations and assumptions inherent in the two techniques, this is a remarkable result. While we have allowed ourselves some freedom to adjust the semi-analytic model in order to maximise the level of agreement with the simulation, it should be noted that in our best fit model, the only changes we have made to the cooling model of Cole \etal (\shortcite{cole2k}) are a slightly smaller core in the gas density profile and an increased cooling time in small halos. Neither of these changes have a large effect on the mean cold gas fraction at $z=0$.

Springel \& Hernquist (\shortcite{s2002}) show that when SPH is formulated in terms of the thermal energy equation, substantial overcooling may occur in halos of fewer than several thousand particles -- for example, gas may cool as it passes through shocks which have been artificially smoothed out by the SPH algorithm. They demonstrate that a new formulation (`entropy SPH') using entropy rather than thermal energy as an independent variable, which conserves both energy and entropy, can significantly reduce this problem. This conclusion would seem to suggest that the quantities of gas cooling in the majority of halos in our SPH simulation may be overestimated. This could explain why a gas profile with a smaller core radius than that used by Cole \etal is required in our semi-analytic model to reproduce the quantities of cold gas in the simulation. However, the Hydra SPH code which we use in this work is significantly different from the GADGET code (\cite{syw2001}) employed by Springel \& Hernquist and it is not clear to what extent our simulation suffers from the overcooling effect.

In an independent investigation carried out concurrently with this one, Yoshida \etal (\shortcite{y2002}) compared gas cooling in SPH simulations carried out using GADGET with a semi-analytic model based on that of Kauffmann \etal (\shortcite{kauffmann99}). This model contains a simpler cooling prescription than used in this work -- the gas within each halo is assumed to trace the dark matter exactly at all times so there is no core radius. Yoshida \etal adopt a similar approach to our own, taking halo merger histories from the dark matter in their SPH simulations and neglecting star formation and feedback in both models. They show results for two of the SPH implementations investigated by Springel \& Hernquist -- one is the entropy SPH implementation discussed above, the other is a `conventional' implementation based on taking the geometric means of the pairwise hydrodynamic forces between neighbouring particles. Yoshida \etal find good agreement between the masses of individual galaxies in their semi-analytic model and the entropy SPH implementation. SPH galaxy masses, however, can differ by a factor of 2 between the two SPH implementations considered, but Yoshida \etal believe the entropy SPH to be the more reliable technique and note that their `conventional' SPH implementation actually suffers the overcooling problem more severely than other conventional implementations, including the Hydra code which we have used here.

Overall, it appears that the differences between cooling rates predicted by SPH and semi-analytic techniques are small, and quite possibly comparable to the uncertainty in the SPH results. As well as providing evidence to support the treatment of cooling in current semi-analytic galaxy formation models, these results show that semi-analytic modelling provides a convenient, alternative way to add a baryonic component to an N-body simulation, which is at least as reliable as an SPH simulation. When used to investigate star formation and feedback prescriptions this approach allows the investigation of large regions of parameter space at little computational cost and so can provide an indication of how these phenomena may be included in full hydrodynamic simulations.

\label{sec:conclusions}
\section*{Acknowledgements}
We acknowledge support from PPARC and the Royal Society.

\end{document}